\documentclass{article}
\usepackage{wrapfig,graphicx}
\newcommand{\Str}{\ensuremath{\mathop{\mathrm {Str}}\nolimits}}
\newcommand{\tr}{\ensuremath{\mathop{\mathrm{tr}}\nolimits}}

\renewcommand{\sinh}{\ensuremath{\mathop{\mathrm{sh}}}}

\newcommand{\mn}{{\mu\nu}}

\newcommand{\fR}{\ensuremath{\mathcal{P}}}

\newcommand{\ve}{\varepsilon}

\newcommand{\cR}{{\cal R}}

\newcommand{\Bk}{\Sigma}

\newcommand{\htt}{\mathbf{T}}
\newcommand{\be}{\begin{equation}}
\newcommand{\ee}{\end{equation}}
\newcommand{\bea}{\begin{eqnarray}}
\newcommand{\eea}{\end{eqnarray}}
\author{D.V. Gal'tsov\thanks{galtsov@grg.phys.msu.su} and V.V. Dyadichev\thanks{vlad@grg1.phys.msu.su}\\\\
\textit{Department of Theoretical Physics,}\\
\textit{Moscow State University, Moscow, Russia}}
\title{%
 \begin{flushright} \begin{small}
  DTP--MSU/03-02 \\ hep-th/0301044
  \end{small} \end{flushright}
\vspace{2.cm}
Non-Abelian brane cosmology}

\begin{document}
\maketitle
\begin{abstract}
We discuss isotropic and homogeneous D-brane-world cosmology with
non-Abelian Born-Infeld (NBI) matter on the brane. In the usual
Fried\-mann-Robertson-Walker (FRW) model the scale non-invariant
NBI matter gives rise to an equation of state which asymptotes to
the string gas equation $p=-\ve/3$ and ensures a start-up of the
cosmological expansion with zero acceleration. We show that the
same state equation in the brane-world setup leads to the Tolman
type evolution as if the conformal symmetry was effectively
restored. This is not precisely so in the NBI model with
symmetrized trace, but the leading term in the expansion law is
still the same. A cosmological sphaleron solution on the D-brane
is presented.
\end{abstract}

In the ``phenomenological'' brane world  approach \cite{RaSu99}
it is assumed that the standard model fields are confined to the
brane by some mechanism whose nature remains unspecified. One
possibility is to invoke the D-brane picture in which vector
fields are generated by open strings with Dirichlet boundary
conditions on the brane. In this case one has to keep in mind
that in order to incorporate the non-Abelian gauge group of the
standard model one has to introduce multiple brane structure with
the number of branes equal to the dimension of the fundamental
representation. These branes may be either coincident or
separated, in the latter case the gauge symmetry will be broken,
with the Higgs vacuum expectation value equal to the brane
displacement. The phase transition is then described as splitting
of an initially coincident set of branes and the subsequent
motion of part of them along the Higgs directions. This opens an
alternative possibility  to view inflation during the phase
transition. The physical universe after the phase transition can
be identified with the subset of coincident branes where the
remaining massless vector fields live.

Here we discuss the homogeneous and isotropic brane cosmology
with the SU(2) Yang-Mills field assuming that the dynamics is
governed by the non-Abelian Born-Infeld action which is an
effective action for  D-branes:
\begin{equation} \label{det} S=\lambda \widetilde{\mathrm{Tr}}\int
\sqrt{-\det(g_{\mu\nu}+F_{\mu\nu}/\beta)}\,d^4 x
 - \kappa^2 \int\, (R_5+2\Lambda_5)\sqrt{-g_{5}}\,d^5x.
\end{equation}
Here $F_{\mu\nu}$ is the matrix-valued field strength (the gauge
group SU(2) being assumed), $g_{\mu\nu}$ is the induced metric on
the brane and $\kappa$, $\Lambda_5$  and $\lambda $ are the 5D
gravitational and cosmological constants and the brane tension
respectively. For generality, the BI critical field strength
parameter $\beta$  is assumed to be different from $\lambda $.
The symbol $\widetilde{\mathrm{Tr}}$ denotes either the ordinary
trace (with an additional prescription to be applied after
evaluating the determinant with the matrix-valued $F$), or
symmetrized trace \Str{} as defined in \cite{Ts97}.

We write the bulk metric as
 \[ ds_5^2=N^2(\tau,y)-a(\tau,y)^2
\left(dr^2+\Bk^2 (d\theta^2+sin^2 \theta d\phi^2)\right) -
  dy^2,
\]where  $\Bk$ is either $\sinh r$, $r$ or $\sin r$  corresponding to
$k = -1,0,1$,  additionally assuming   that it is invariant under
reflection $y\to -y$. The induced metric on the brane is given by
the first two terms with the limiting values of $N$ and $a$ at
$y=0$.
The gauge field exists only on the brane and is given by
$G_6$-invariant ansatz found in \cite{GaVo91}, it is
parameterized by a single function $w$  of time:
\begin{eqnarray}\label{Fform}
\mathcal{F}=\dot{w}\left(\htt_r\,dt\wedge dr
     + \htt_\theta \Bk\,dt \wedge d\theta
     + \htt_\phi \Bk \sin \theta \,dt \wedge d\phi \right) + \nonumber\\
+ \Bk(w^2-k)\left(\htt_\phi\,dr \wedge d\theta
     - \htt_\theta \sin \theta\,dr \wedge d\phi
     + \htt_r \Bk  \sin \theta\,d\theta \wedge d\phi \right),
\end{eqnarray}
where $\htt_r$, $\htt_\theta$ and $\htt_\phi$ is the gauge group
generators (normalized as $\tr \htt_i\htt_j=2\delta_{ij}$) projected
onto the spherical legs.

The computation of the reduced action is straightforward for the
``ordinary trace'' prescription \cite{DyGaZoZo01}. For the
symmetrized trace one has to expand the square root into powers
of the matrix-valued $F_{\mu\nu}$, to perform symmetrization in
the generator products, to take the trace, and then to attempt a
resummation. In the present case this calculation is simplified by
noting that due to spherical symmetry the generators enter only
through the combination $\htt_r^2+\htt_\theta^2+\htt_\phi^2$. It
is easy to check that for the n-th power of this quantity one has
\[
\Str \left(\htt_r^2+\htt_\theta^2+\htt_\phi^2\right)^n = 2 (2 n +
1).
\]
Using this relation the resummation  of the lagrangian can be
done in the closed form and we arrive at the following
one-dimensional action
\begin{equation}\label{stractn}
S^{Str}_1=-\lambda \int \; dt N a^3
\frac{1-2K^2+2V^2-3V^2K^2}{\sqrt{1- K^2+ V^2-  K^2V^2}}.
\end{equation}
For the ordinary trace the one-dimensional action reads
\cite{DyGaZoZo01}:
\begin{equation}\label{tractn}
S^{tr}_1=-\lambda \int \; dt N a^3 \sqrt{1-3 K^2+3 V^2- 9 K^2V^2}.
\end{equation}
In these formulas
\[
K^2=\frac{\dot{w}^2}{\beta^2 a^2 N^2},\qquad
V^2=\frac{(w^2-k)^2}{\beta^2 a^4},
\]
and functions $N$ and $a$ refer to the metric on the brane.

Varying the action (\ref{det}) with respect to the induced
metric one finds the following energy-momentum tensor of the gauge
field
\begin{equation}
    T_{\mu\nu} =\lambda  \widetilde{\mathrm{Tr}} (\cR^{-1}(g_\mn+\beta^{-2}(\frac 12 F^{\lambda\rho}F_{\lambda\rho} g_\mn
    -F_{\mu\lambda}F_{\nu}\,{}^{\lambda}))- \lambda g_\mn,
\end{equation}
where
\[
\cR=\sqrt{-\det(g_{\mu\nu}+F_{\mu\nu}/\beta)},
\]
which is not traceless as it is in the usual Yang-Mills theory. In
the context of the FRW cosmology this means that the equation of
state will be different from $p=\ve/3$ found in the latter case
for isotropic and homogeneous configurations \cite{GaVo91}. In
the model with the ordinary trace (\ref{tractn}) the energy
density is \cite{DyGaZoZo01}
\begin{equation}\label{epsdef}
    \ve= T ^0_0=\lambda   \left(\fR -1\right),\quad
    \fR=\sqrt{\frac{1+V^2}{1-K^2}},
\end{equation}
while the pressure is given by
\begin{equation}\label{pressdef}
     p=\frac13 \lambda  \left(3-\fR
       -2\fR^{-1}\right).
\end{equation}
Comparing (\ref{epsdef}) and (\ref{pressdef}) one finds the state
equation
\begin{equation}\label{state}
     p=\frac{\ve}{3}\frac{(\lambda-\ve )}{(\lambda+ \ve)},
\end{equation}
where the  brane tension $\lambda$  plays the role of the
critical energy density at which the pressure vanishes; for
larger energies pressure becomes negative and attains the
limiting value $p=-\ve/3$ at high density. This is the string gas
state equation which apparently reflects the string origin of the
effective action (\ref{det}). In the opposite limit $\ve\ll
\lambda $ one finds the hot matter state equation $p=\ve/3$, in
fact in this limit the scale invariance of the action is
restored.  The dynamics of the gauge field $w$ in the usual FRW
model is then expressible in terms of the Jacobi elliptic
functions \cite{DyGaZoZo01}. Remarkably, the space-time evolution
can be separated from that of the gauge field. Namely, using the
conservation equation
\begin{equation}\label{conserv}
\dot{\ve}+3(\ve + p)\frac{\dot{a}}{a}=0,
\end{equation}
one can find the following dependence of energy density on the
scale factor $a$:
\begin{equation}\label{falloff}
    \ve = \lambda  \left(\sqrt{1+\frac{C}{a^4}}-1\right),
\end{equation}
where $C$ is the integration constant. This   law interpolates
between $a^{-4}$ dependence (arising for conformal fields) in the
limit of small energy densities and $a^{-2}$ dependence at high
density, reflecting the state equation $p=-\ve/3$.

Let us turn to the brane world setup using the formulation of
\cite{Bi99,Bi99-0}. The brane analogue of the constraint equation
reads
\begin{equation}\label{RSaeq}
    \left(\frac{\dot{a}}{a}\right)^2=\frac{\kappa^2}{6} \Lambda
  + \frac{\kappa^4}{36}
  (\lambda  +\ve)^2+\frac{\mathcal{E}}{a^4}-\frac{k}{a^2},
\end{equation}
where $\mathcal{E}$ is integration constant corresponding to the
bulk Weyl tensor projection (``dark radiation'', see
\cite{ShMaSa99}), and $\ve$ is the energy density on the brane
defined by (\ref{epsdef}). We additionally fix the gauge by
setting $N(\tau,0)=1$, thus identifying $\tau$ with the proper
time on the brane. The constant terms in this equation, being
grouped together, give the value of an effective 4D cosmological
constant $\Lambda_4$ and the
 4D  Newton constant:
\begin{equation}\label{lam4def}
  \Lambda_4 = \frac12 \kappa^2(\Lambda+\frac16 \kappa^2 \lambda  ^2),
  \quad  G_{(4)}=\frac{\kappa^4 \lambda  }{48 \pi }.
\end{equation}
Note that the conservation equation (\ref{conserv}) remains valid
in the brane scenario too, and consequently the energy density
falloff law (\ref{falloff}) still holds. Most surprisingly, the
equation (\ref{RSaeq}) simplifies greatly when the dependence Eq.
(\ref{falloff}) is inserted, and one obtains
\begin{equation}\label{BIRS}
  \left(\frac{\dot{a}}{a}\right)^2=\frac{8\pi G_{(4)}}{3}\Lambda_4
  +\frac{\mathcal{C}~}{a^4}-\frac{k}{a^2},
\end{equation}
where  the constant $\mathcal {C}= \mathcal{E} + \kappa^4\lambda^2
C/36$ includes contributions from both the ``dark radiation'' and
the energy density of the NBI matter. This equation fully
coincides with the corresponding equation for the ``hot'' FRW
cosmology in the non-brane scenario \cite{GaVo91},  and therefore
we are led to the standard Tolman solutions. In particular, this
means that near the singularity the scale factor grows like
$t^\frac12$.

There is, however, one essential distinction from the
conventional hot cosmology: since the r.h.s. of the Eq.
(\ref{BIRS}) includes the ``dark radiation'' contribution, now the
total effective energy density need not be positive, i.e. the
constant  $\mathcal C$ is not a priori of a definite sign. This
gives rise to  solutions that could not arise in the conventional
hot cosmology.  One is the static flat (in the 4D sense) solution
with the vanishing cosmological constant
 and nonzero energy density:
\[\dot{a}=0, \qquad\mbox{and after simple scaling}\qquad g_\mn=\eta_\mn.
\]
It can be obtained from the Eq. (\ref{BIRS}) with  the following
parameters: $\Lambda_4=0$, $\mathcal{C}=0$, $k=0$. The
possibility of such a solution in the brane-world setup was
investigated in \cite{GeMa01} assuming the fluid matter content.
It was found that the necessary state equation has the general
form coinciding with our NBI equation (\ref{state}). Therefore,
the ordinary trace  NBI model on the brane provides a very natural
realization of the static scenario.

Another new interesting possibility is the avoidance of
cosmological singularity without the cosmological constant. Such
solutions can be obtained from (\ref{BIRS}) setting
$\Lambda_4=0$, $\mathcal{C}<0$, $k=-1$. This corresponds to the
open universe. The solution reads:
\begin{equation}\label{nonsingtr}
a  = \sqrt{|\mathcal{C}|+(t-t_0)^2},
\end{equation}
where $t_0$ is the integration constant. The universe contracts
to the minimal radius $a  = \sqrt{|\mathcal{C}|}$, bounces and
reexpands.

\begin{wrapfigure}[13]{r}[2pt]{6.2cm}
    \begin{minipage}{6cm}
     \vspace{-5pt}

        \includegraphics[width=5cm]{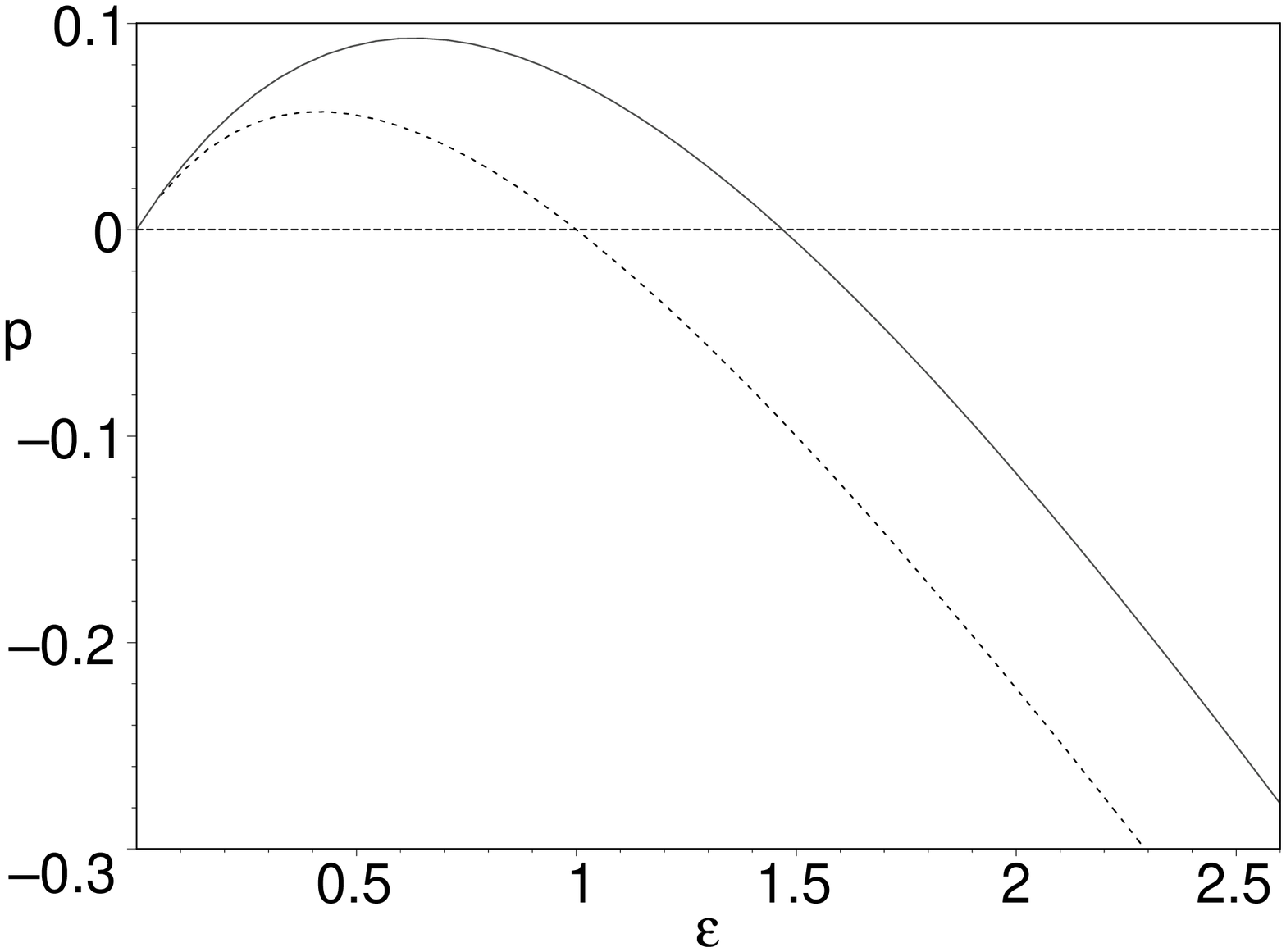}

	\begin{centering}
            \small Effective state equation for the Str
            (solid line) and tr (dotted line)
            cosmological sphaleron ($w\equiv 0$) models.
        \end{centering}
    \end{minipage}
\end{wrapfigure}
The situation is qualitatively similar, though not precisely the
same in the model with symmetrized trace (\ref{stractn}). One has
to solve the full system of equations consisting of one derived
from the action (\ref{stractn}) for the gauge field and the
equation (\ref{RSaeq}) for the metric. In this case one can not
extract an equation of state since the pressure is no more
expressible as a function of the energy only. However this can be
done in the limiting cases of low and high densities, or on the
particular classes of solutions. In the limiting cases one has
the same equations of state as in the ordinary trace model
\begin{eqnarray}
    p &\approx& -\frac 13 \ve,  \qquad \mbox{ as \quad $\ve \to \infty$,}\\
    p &\approx& \frac 13 \ve,  \;\;\; \qquad \mbox{ as \quad $\ve \to
    0 $.}
\end{eqnarray}
This   indicates that the behavior of the solutions near the
singularity (if it exists)   also remains of the  Tolman type. At
late times the behaviour is again typical for the radiation
dominated universes. So in the symmetrized trace model the main
qualitative features of the hot universe pertain.

There are also some ``exotic'' solutions with negative values of
dark energy contribution $\mathcal{E}$. In the case of the
vanishing cosmological constant $\Lambda_4$ the static Einstein
universe cannot exist. This is because the state equation
(\ref{state}) needed for the static universe is fulfilled in the
symmetrized trace model only asymptotically. Instead, if the value
of the dark radiation constant $\mathcal{E}$ is below some
negative critical limit, there appears solutions starting from
the initial singularity, expanding to a finite $a$ and then
collapsing back to the singularity. The analogue of the
nonsingular solution (\ref{nonsingtr}) with $k=-1$ exists in the
symmetrized trace model too.

One particular solution is worth to be mentioned. It can be
checked that $w\equiv 0$ is a non-trivial solution to the field
equations for $k=\pm 1$ like in the ordinary YM theory
\cite{GiSt95}. In the closed case this is the cosmological
sphaleron sitting at the top of the potential barrier separating
topologically distinct vacua. The scale factor starts with an
expansion
\begin{equation}
a=(4\lambda^2+ \mathcal{E}\beta^2)^{1/4}\sqrt{\frac{2t}{\beta}}
-\frac{k\beta^2}{2} (4\lambda^2+ \mathcal{E}\beta^2)^{-1/4}
\left(\frac{2t}{\beta}\right)^{3/2} + O (t^{5/2}),
\end{equation}
the global behavior of the solution being similar to that in the
non-brane scenario with the ordinary YM matter \cite{GiSt95}. The
effective equation of state curves are plotted above for the
symmetrized and the ordinary trace models.

One of the authors (DVG) would like to thank the Organizing
Committee for support to attend the Conference. The work was
supported by the RFBR under grant 00-02-16306.

\end{document}